\documentclass[conference]{IEEEtran}
\IEEEoverridecommandlockouts

\usepackage{cite}
\usepackage{amsmath,amssymb,amsfonts}
\usepackage{algorithmic}
\usepackage{graphicx}
\usepackage{textcomp}
\usepackage{tikz}
\usepackage{pgfplots}
\usepackage{xcolor}
\usepackage{soul,color}
\usepackage{makecell}
\usepackage[font=small,skip=2pt]{caption}
\usepackage{adjustbox}
\usepackage{hyperref}
\usepackage{array}
\usepackage{paralist}
\usepackage{booktabs}
\usepackage{siunitx}
\usepackage{tablefootnote}
\usepackage{subfigure}
\usepackage{float}
\DeclareSIUnit\ms{ms}
\def\BibTeX{{\rm B\kern-.05em{\sc i\kern-.025em b}\kern-.08em
    T\kern-.1667em\lower.7ex\hbox{E}\kern-.125emX}}

\usepackage[acronyms,nomain,xindy,nowarn]{glossaries}
\makeglossaries
\newacronym{5g}{5G}{Fifth-Generation Wireless Systems}
\newacronym{ai}{AI}{artificial intelligence}
\newacronym{ann}{ANN}{Artificial Neural Network}
\newacronym{ask}{ASK}{amplitude-shift keying}
\newacronym[plural={AWGN}, \glsshortpluralkey={AWGN}]{awgn}{AWGN}{additive white Gaussian noise}
\newacronym{au}{AU}{aerial user}
\newacronym{ap}{AP}{access point}
\newacronym{bec}{BEC}{binary erasure channel}
\newacronym{ber}{BER}{bit error rate}
\newacronym{bler}{BLER}{block error rate}
\newacronym{bpsk}{BPSK}{binary phase-shift keying}
\newacronym{bs}{BS}{base station}
\newacronym{bbu}{BBU}{baseband unit}

\newacronym{cdf}{CDF}{cumulative distribution function}
\newacronym{cnn}{CNN}{convolutional neural network}
\newacronym{csi}{CSI}{channel state information}
\newacronym{comp}{CoMP}{coordinated multi-point}
\newacronym{cran}{C-RAN}{cloud-radio access network}
\newacronym{cca}{CCA}{cloud coverage area}

\newacronym{dnn}{DNN}{deep neural network}
\newacronym{drl}{DRL}{deep reinforcement learning}
\newacronym{dqn}{DQN}{deep Q-network}
\newacronym{daps}{DAPS}{dual active protocol stack}
\newacronym{ecc}{ECC}{error-correcting code}
\newacronym{ecdf}{ECDF}{empirical distribution function}
\newacronym{ee}{EE}{energy efficiency}
\newacronym{ec}{EC}{edge cloud}
\newacronym{cc}{CC}{central cloud}
\newacronym{e2e}{E2E}{end-to-end}

\newacronym{ff}{FF}{feed-forward}
\newacronym{ffnn}{FF-NN}{feed-forward neural network}
\newacronym{fsk}{FSK}{frequency-shift keying}
\newacronym{fs}{FS}{functional split}
\newacronym{fsm}{FSM}{finite state machine}

\newacronym{gm}{GM}{Gaussian mixture}
\newacronym{GOPS}{GOPS}{Giga Operation Per Second}

\newacronym{hmappo}{H-MAPPO}{hierarchical multi-agent proximal policy optimization}
\newacronym{hdrl}{HDRL}{Hierarchical deep reinforcement learning}
\newacronym{hmarl}{HMARL}{Hierarchical Multi-Agent Reinforcement Learning}

\newacronym{iid}{i.i.d.}{independent and identically distributed}
\newacronym{il}{IL}{information leakage}
\newacronym{iot}{IoT}{internet of things}
\newacronym{ilp}{ILP}{Integer Linear Programming}

\newacronym{lb}{LB}{lower bound}
\newacronym{los}{LoS}{line-of-sight}
\newacronym{lstm}{LSTM}{long short-term memory}

\newacronym{mac}{MAC}{multiple access channel}
\newacronym{map}{MAP}{maximum a posteriori}
\newacronym{mc}{MC}{Monte Carlo}
\newacronym{mdp}{MDP}{Markov decision process}
\newacronym{mimo}{MIMO}{multiple-input multiple-output}
\newacronym{miso}{MISO}{multiple-input single-output}
\newacronym{ml}{ML}{machine learning}
\newacronym{mmwave}{mmWave}{millimeter wave}
\newacronym{mop}{MOP}{multi-objective programming problem}
\newacronym{mrc}{MRC}{maximum ratio combining}
\newacronym{msc}{MSC}{modulation and coding scheme}
\newacronym{mse}{MSE}{mean squared error}
\newacronym{msac}{MSAC}{masked soft actor-critic}
\newacronym{mappo}{MAPPO}{multi-agent proximal policy optimization}
\newacronym{madrl}{MADRL}{multi-agent deep reinforment learning}
\newacronym{milp}{MILP}{Mixed Integer Linear Programming}
\newacronym{mec}{MEC}{Multi-access Edge Computing}

\newacronym{nlos}{NLoS}{non-line-of-sight}
\newacronym{nn}{NN}{neural network}
\newacronym{nve}{NVE}{normalized validation error}
\newacronym{ric}{Near-RT RIC}{Near-Real Time RAN Intelligent Controller}
\newacronym{noma}{NOMA}{Non-orthogonal multiple access}
\newacronym{nfv}{NFV}{network function virtualization}

\newacronym{oru}{O-RU}{O-RAN radio unit}
\newacronym{oran}{O-RAN}{open-radio access network}
\newacronym{odu}{O-DU}{O-RAN distributed unit}

\newacronym{pdf}{PDF}{probability density function}
\newacronym{pwc}{PWC}{polar wiretap code}
\newacronym{ppo}{PPO}{proximal policy optimization}
\newacronym{pf}{PF}{processing function}
\newacronym{qam}{QAM}{quadrature amplitude modulation}
\newacronym{qos}{QoS}{quality of service}

\newacronym{ra}{RA}{rearrangement algorithm}
\newacronym{rbf}{RBF}{radial basis function}
\newacronym{relu}{ReLU}{rectified linear unit}
\newacronym{ris}{RIS}{reconfigurable intelligent surface}
\newacronym{rl}{RL}{reinforcement learning}
\newacronym{rnn}{RNN}{recurrent neural network}
\newacronym{rf}{RF}{random forest}

\newacronym{sac}{SAC}{soft actor-critic}
\newacronym{sgd}{SGD}{stochastic gradient descent}
\newacronym{sgp}{SGP}{secure goodput}
\newacronym{simo}{SIMO}{single-input multiple-output}
\newacronym{sinr}{SINR}{signal-to-interference-plus-noise ratio}
\newacronym{siso}{SISO}{single-input single-output}
\newacronym{snr}{SNR}{signal-to-noise ratio}
\newacronym{svm}{SVM}{support vector machine}

\newacronym{tvd}{TVD}{total variation distance}
\newacronym{ttr}{TTR}{time to trigger}

\newacronym{uav}{UAV}{unmanned aerial vehicle}
\newacronym{ub}{UB}{upper bound}
\newacronym{urllc}{URLLC}{ultra-reliable low latency communication}

\newacronym{v2v}{V2V}{vehicle-to-vehicle}
\newacronym{v2x}{V2X}{vehicle-to-everything}
\newacronym{vm}{VM}{virtual machine}
\newacronym{vnf}{VNF}{virtual network function}

\newacronym{zoc}{ZOC}{zero-outage capacity}
\setacronymstyle{long-short}
\glsdisablehyper
\usepackage{todonotes}

\usepackage{physics}

\def\BibTeX{{\rm B\kern-.05em{\sc i\kern-.025em b}\kern-.08em
    T\kern-.1667em\lower.7ex\hbox{E}\kern-.125emX}}

\title{Surfing the SWAVES: Lifecycle-aware Service Placement in MEC}

\author{\IEEEauthorblockN{Federico Giarrè, Holger Karl}
\IEEEauthorblockA{\textit{Hasso-Plattner Institute (HPI)} \\
Email: federico.giarre, holger.karl at hpi.de}}
\begin{document}

\maketitle

\begin{abstract}
In \gls{mec} networks, users covered by a mobile network can exploit \glspl{ec}, computational resources located at the network's edge, to execute \glspl{vnf}. \glspl{ec} are particularly useful when deploying \glspl{vnf} with strict delay and availability requirements.
   % as in the case of \glspl{vnf} for \gls{urllc}-based services.
  %e.g., for  \gls{urllc}-based services.
As users roam in the network and get handed over between cells, deployed \glspl{vnf} must follow users to retain the benefits of edge computing.
Yet, having \glspl{vnf} ready at the closest EC can be challenging:
\begin{inparaenum}[(i)]
    \item \glspl{ec} are not usually powerful enough to store and run any combination of \glspl{vnf} simultaneously; 
    \item if a \gls{vnf} is not available at the needed \gls{ec}, a series of time-consuming operations has to be performed before the \gls{vnf} becomes operational.
\end{inparaenum}
%As service requirements become stricter, careful planning of service deployment, subject to the constraints described, becomes paramount to ensure that such constraints are met.  
These limitations can be addressed by proactively starting \glspl{vnf} instances at (likely) future locations, balancing better latency properties against higher resource usage. Such proactive deployment does need forecasting of user movements, but these will be imperfect, creating yet another tradeoff.
We present our approach to this service provisioning problem, SWAVES.
When compared on the ratio of users' unsuccessful packets, SWAVES improves such metric by orders of magnitude with respect to other proposed heuristic.
\end{abstract}

%\fmhkn{two thoughts: SWAVES is not a forecasting approach as such, is it? It makes use of forecasts, but does not produce them itself. Optimality gap: In the optimzation community, this has a strcit meaning (the relative difference between an upper and a lower bound on the optimal solution; hence the worst-case deviation from the optimum). Is that really what you mean? IIRC, you are comparing solutions against an oracle, so yes, has the bound-distance-from-optimum aspect, but it still might be misunderstood? how about: ``In this study, we introduce SWAVES and compare its performance to both an oracles and plausible baselines. SWAVES' difference to the oracle is orders of magnitudes smaller than that of all  baselines.''Minor point: actions as predicates! ``showed a reduction'' is ``reduces'' ! and ``baseline'' is such fancy ML lingo :-)  }


\section{Introduction}

We consider a scenario where mobile users access services provided by a \gls{mec} infrastructure, with maximum \gls{e2e} latency in the milliseconds and high availability requirements. These services are deployed via \glspl{vnf} running in computing premises situated close to \gls{bs}, i.e. \glspl{ec}.
In \gls{mec}, users can deploy \glspl{vnf} at \glspl{ec} to comply with their \gls{qos} requirements thanks to proximity, reducing latency.

\begin{figure}[h]
    \includegraphics[width=\linewidth]{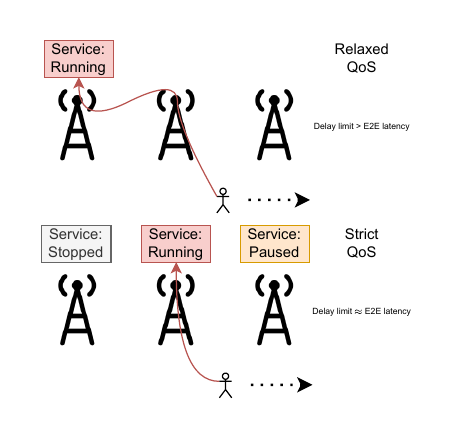}
    \caption{If forwarding requests across the network does not violate QoS constraints (upper figure), services can be deployed at the most convenient premise. When strict QoS requirements have to be met (lower figure), services must follow the users in their movements.}
    \label{fig:swawesEdgeCases}
  \end{figure}

In scenarios where \gls{qos} requirements are relaxed, \gls{vnf} placement can be flexible and optimized for resource usage and load balancing, even if distant from the user, as shown is \autoref{fig:swawesEdgeCases}.
But what if latency requirements become tight and users are mobile? 
\glspl{vnf} deployment need to follow users' mobility to maintain the benefits of \gls{mec}.
As users connect to new \glspl{ec}, however, the needed \gls{vnf} may not be available yet; there might not even be the software for it. One solution is to reroute user traffic to a running instance of the \gls{vnf} in another location, increasing  experienced delay \cite{ouyang_follow_2018} while starting the \gls{vnf} at the new location. But this approach easily violates strict \gls{qos} requirements of certain classes of services. 
%Conversely, services with strict latency requirements could be deployed along or close to user paths to retain proximity and benefit from \gls{mec}, as illustrated in \autoref{fig:swawesEdgeCases}.
Another approach could be to have every \gls{vnf} always running at every \gls{ec} but, by definition, \glspl{ec} have constrained computational resources, rendering this option unfeasible.
To support such services, it is necessary to pre-emptively start \glspl{vnf} instances along user paths to retain proximity and benefit from \gls{mec}. But to start a \gls{vnf}, a sequence of time-consuming operations has to be performed, known as service \emph{lifecycle} \cite{stahlbock_optimization_2022,he_virtual_2020}: from the retrieval and building of the executable if not available, to the start of a \gls{vm} or container if virtualized, and finally responding to requests.
These operations need variable time depending on the type of service, executable type, and network conditions, delaying the effectiveness of service provisioning decisions up to tens of seconds. 

%We study how service lifecycle affects provisioning in \gls{mec}, a topic yet to be addressed by the \gls{nfv} and \gls{mec} communities.

% \begin{figure}
%     \includegraphics[width=\linewidth]{img/firstpage.pdf}
%     \caption{If forwarding requests across the network does not violate QoS constraints (upper figure), services can be deployed at the most convenient premise. When strict QoS requirements have to be met (lower figure), services must follow the users in their movements.}
%     \label{fig:swawesEdgeCases}
%   \end{figure}

% \begin{figure}
%     \centering
%     \includegraphics[scale=0.6]{img/heatmap.pdf}
%     \caption{Likelihood of service being needed in cells in the near future.}
%     \label{fig:heatmap}
% \end{figure}

\begin{figure}[h]
    \centering
    \includegraphics[scale=0.7]{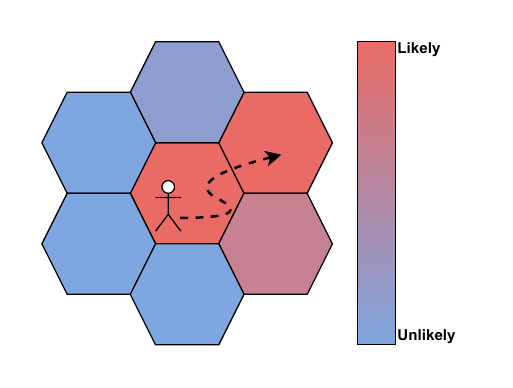}
    \caption{Heatmap of service being needed in cells in the near future.}
    \label{fig:heatmap}
\end{figure}

To address service lifecycle during provisioning along users' path, exact knowledge about future users' connections are needed. However, such knowledge is not obtainable in the real world. A key helper function to deal with this problem is forecasting of user behavior, in particular location \cite{ghosh_mobi_iost}.
This can be exploited to pre-emptively deploy users' required \glspl{vnf}  \emph{in time} and \emph{at the right location}~\cite{yu_characterizing_2020}. 
On the other hand, we cannot expect forecasts to be always accurate, especially as the randomness in user movements increases~\cite{ouyang_follow_2018}.  
To tackle this problem, instead of forecasting the exact next location of users, we can look at the likelihood of a service being needed in a cell further in the future (as shown in \autoref{fig:heatmap}).
These forecasts can be finally used to commence \glspl{vnf}' startup operations where needed, preparing them where they will be (likely) needed and, most importantly, removing them where they are not useful anymore. 
From a logical viewpoint, a service spreads to nearby \glspl{ec} starting from users requesting it, creating a \emph{software wave}, hence the name of our approach, SWAVES.

Our contributions in this paper are the following:

\begin{itemize}
    \item We characterize service lifecycle via a \gls{fsm}, describing the sequence of states a generic \gls{vnf} has to go through and the time it takes before being operational. We then consider such \gls{fsm} when placing  services, including their installation, startup and usage.
    \item We introduce SWAVES, a lifecycle-aware service provisioning approach, robust to the randomness of user movement.
    \item We experimentally validate SWAVES with respect to optimal and heuristic solvers under different \gls{qos} strictness and mobility randomness scenarios.
\end{itemize}

\section{Related Work}
The idea of services following users in a network has been thoroughly explored by the networking, \gls{mec} and \gls{nfv} communities, but many aspects of service migration are often not considered.
Two different studies \cite{ouyang_follow_2018,wang_dynamic_2015} explore a concept similar to this work with roaming users in a \gls{mec} network. While these studies optimally solve the service placement, they do not consider resource constraints on \glspl{ec} and the time taken to complete the service migration.
Similarly, in \cite{taleb_follow-me_2019}, Taleb \emph{et al.} explore the problem from a higher-level perspective, moving from the \gls{mec} architecture to a geo-distributed cloud network. Here, the downtime caused by service migration is taken into consideration. As per the previously discussed studies, constrained computational resources are not considered.
%Above studies have user-cell association described by a Markov chain. This is done to estimate the distance between the user and the serving \gls{ec}, defining then a policy on service migration or user handover. However, such Markov chains can only represent ideal topologies with strong assumptions.
Liu \emph{et al.}  \cite{liu_data_2022} describe a lower-level approach to the problem, where the focus is reducing the overhead for  service migration. Here, a \gls{dnn} is used to optimally place services. Yet, finite resource of \glspl{ec} are not considered in this work. Similary, in \cite{CHEN2023108552} Chen \emph{et al.} employ \gls{drl} and user movement predictions to migrate services,   minimizing packet loss and service disruption during roaming. The study accounts for finite resources but does not consider service lifecycle as part of the placement problem.
The tradeoff between \gls{qos} and the cost of deployment is explored in \cite{mada_latency-aware_2020,he_virtual_2020} where the authors optimize placement with respect to limited computational and network resources. While not a comprehensive characterization of a service lifecycle, \cite{he_virtual_2020} acknowledges that this step is missing from other service placement studies.   

% A summary of the discussed studies and how they compare to this work can be found in  \autoref{tab:rw}.
\autoref{tab:rw} provides a summary of the related work discussed. Multiple studies addressed optimal service placement, but to our knowledge, none characterize and consider the full lifecycle of a   service as part of the placement problem

\begin{table*}[h!]
    \centering
    \caption{Literature comparison}
    \begin{tabular}{p{0.25\linewidth}|cccc}
        \toprule
        Study & Finite Resources & Service Migration & Forecasting & Service Lifecycle \\
        \midrule
        Ouyang \emph{et al.} \cite{ouyang_follow_2018} & x & $\checkmark$ &  $\checkmark$ & x \\
        Wang \emph{et al.} \cite{wang_dynamic_2015} &  $\checkmark$ &  $\checkmark$ & x & x \\
        Taleb \emph{et al.} \cite{taleb_follow-me_2019} & x &  $\checkmark$ & x & x \\
        Liu \emph{et al.} \cite{liu_data_2022} &  x&$\checkmark$ & $\checkmark$ & x \\
        Chen \emph{et al.} \cite{CHEN2023108552} &  $\checkmark$&$\checkmark$ & $\checkmark$ & x \\
        He \emph{et al.} \cite{he_virtual_2020} &  $\checkmark$&$\checkmark$ & $\checkmark$ & x\footnotemark \\
        Mada \emph{et al.} \cite{mada_latency-aware_2020} &  $\checkmark$ & $\checkmark$ & $\checkmark$ & x \\
        %Ghosh \emph{et al.}\cite{ghosh_mobi-iost_2020} & x &  $\checkmark$ &  $\checkmark$ & x \\
        This study &  $\checkmark$ & $\checkmark$ & $\checkmark$ & $\checkmark$ \\
        \bottomrule
    \end{tabular}
    \label{tab:rw}
    
  \end{table*}
  
\section{System Model}
Let $\mathcal{U}$ be the set of users moving in a network composed of a set of \glspl{bs} $\mathcal{B}$. Each \gls{bs} is extended by an \gls{ec}, with an array of available resources $R_{\text{tot}}$ and connected to its \gls{bs} via a wired link. At all times, each user $u$ is connected to one \gls{bs} $b_u(t)$ and one \gls{ec} $e_u(t)$.

We also assume \gls{daps} to be enabled for handovers. \glspl{daps} allows users to maintain their connection with their old \gls{bs} while establishing a new one with another. This avoids service interruptions generated during handovers, essential to comply with stringent \gls{qos} requirements \cite{lee_intelligent_2022}

\subsection{Infrastructure}
\begin{figure}
    \centering
    \includegraphics[width=\linewidth]{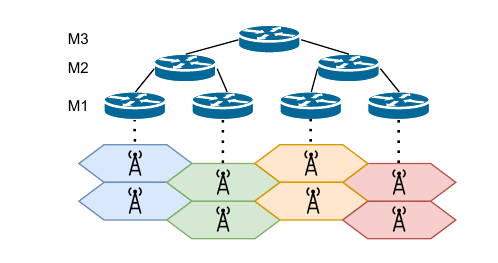}
    \caption{Logical topology considered for the network.}
    \label{fig:topo}
\end{figure}

%Services  used in \gls{mec} can differ in many aspects, one of which is the time taken to process and serve users. To retain generality, we only consider \gls{e2e} delay.\fmhkn{? not sure I got how these two sentences relate to each other?}
We consider a tree-shaped logical topology as shown in \autoref{fig:topo} \cite{coll-perales_end--end_2022}, with \glspl{bs} connected via multiple layers of multiplexing points $M_1, M_2, \dots$
The \gls{e2e} delay between users and the \gls{ec} serving them, modeled after \cite{firouzi_delay-sensitive_2024}. 
Here, the delay is calculated by adding the delays generated by links users' requests go through. As in \cite{firouzi_delay-sensitive_2024}, we consider users, \glspl{bs}, and \glspl{ec} to be close enough to deem propagation delay negligible. We consider transmission delay can to be negligible \cite{firouzi_delay-sensitive_2024}, as in \cite{firouzi_delay-sensitive_2024}. Hence,
we account only for processing and queuing delays. Processing delay is described as the time taken to process a packet when receiving and sending and, as in \cite{coll-perales_end--end_2022}, it is considered a constant $t_p$ .

As in \cite{firouzi_delay-sensitive_2024}, queuing delay for the wireless medium is treated as a $M/M/1$ queue, while for wired links as $M/D/1$ queues. User $u$  makes requests at  rate $\lambda_u$; each link $l$ has a service rate $\mu_l$ and requests arrive at it with rate $\lambda_l$, which is (by Little's law) the sum of the arrival rates of all users using that link. 
% . Using Little's law and envisioning the system as a sequence of Markovian queues,\fmhkn{? the M/D queues are not Markovian? but for the sum of arrival rates, they don't need to be Markovian, so the comment is confusing?}, the arrival rate on a particular link, i.e. $\lambda_l$, consists of the sum of the arrival rates of users using that link.
While $\mu_l$ is constant for cabled links and, as a $M/D/1$ queue, so is the service time per packet,  it is bounded by Shannon theorem's for the wireless link \cite{liu_data_2022,firouzi_delay-sensitive_2024}. We consider each intermediate hop between users and \glspl{ec} to both send and receive the packets sent by users \cite{coll-perales_end--end_2022}. Conversely, users and \glspl{ec} each either receive or send packets. 

Finally, given $\mathcal{L}$ the set of wired links to the \gls{ec} hosting the service and 
 considering $\rho_l$ the processing rate of a link, we compute the total \gls{e2e} delay perceived by user $u$ as:
\begin{equation}\label{eq:delay}
    \underbrace{\frac{1}{\mu_u - \lambda_u}}_{\text{wireless link queuing}} + \underbrace{2(|\mathcal{L}|+1)t_p}_{\text{processing time}}+\underbrace{\sum^{\mathcal{L}}_l\frac{1}{\mu_l}+\frac{\rho_l}{2\mu_l (1- \rho_l)}  }_{\text{queuing per wired link}}.
  \end{equation}

\footnotetext{The study considers starting time for services, but does not provide nor use a model to describe the entire lifecycle.}

Note that the path connecting any user with any \gls{ec} is always composed of at least the wireless medium, one cabled link and additional cabled links for forwarding. 

% \begin{figure}
%     \centering
%     \includegraphics[width=\linewidth]{img/delay.pdf}   
%     \caption{Network as a chain of $M/M/1$ queues \hl{fix image, bs are no longer connected directly}}
%     \label{fig:delay}
% \end{figure}
%
\
\subsection{Service Model}
We consider a set $\mathcal{V}$ of multi-tenant \glspl{vnf}; each instance can serve multiple users across the network. Each instance $v$ requires resources $R_v$ to run.
$R_v$ is a vector of fixed values, one per type of resource (e.g., CPU, memory, disk),
and is enough for the services to comply with the required service level agreement (SLA)\footnote{More general resource requirements models, e.g., having required resources as a linear function of offered load, can be easily integrated in our model, but it is outside the scope of this paper. }.

Stateless \glspl{vnf} can easily work in mobility scenarios since users can instantly be attached to a new instance without further operations. On the other hand, stateful \glspl{vnf} need to migrate user context from a previous  instance to correctly serve the user. Migrating such context requires time, depending on the context size $V_{\text{mem}}$.
%Intuitively, $V_{\text{mem}}$ can differ from service to service, but .
For simplicity, we assume $V_{\text{mem}}$ to be constant even though different services may have extremely different  $V_{\text{mem}}$. 

The duration of the context migration can be computed similarly to \gls{e2e} delay, which is dominated by the queuing and processing time of the hops to traverse the infrastructure. Conversely from the delay model presented in \eqref{eq:delay}, however, the size of data to transmit is not negligible, so the transmission delay is also accounted for.

%
%\subsection{Service Migration}
%If in the case of stateless service we can reasonably assume that as soon as the service is ready at the new \gls{ec} the user can connect to the new service, this is not true for the stateful services. 
%When it comes to stateful services, a common technique for service migration is live migration \cite{liu_performance_2011}. This is composed of two phases: \emph{Pre-Copy} and \emph{Stop-and-Copy}. In the Pre-Copy phase, the memory of the current service is copied to the service in the new location. This process is repeated until a specific condition is met (i.e., number of rounds), to copy back memory pages dirtied in the meanwhile. In the Stop-and-Copy phase, both the old and new services are paused to copy the remaining dirty pages. As the second phase is the only part in which the service is not available at any of location, the downtime created by the live migration is based on the duration of the Stop-and-Copy phase.
%In particular, as per \cite{strunk_costs_2012,liu_performance_2011}, given the working set size $V_{mem}$ of a service, the page dirtying rate $\rho$, and a constant time for resuming the service $T_{\text{resume}}$, the Stop-and-Copy phase can be computed as:

\begin{equation}
    %T_{s\&c} = \frac{\rho V_{mem}}{B}.
    \underbrace{\sum^\mathcal{L}_l\frac{V_{\text{mem}}}{\mu_l}}_{\text{transmission delay}}+ \underbrace{2(|\mathcal{L}|+1)t_p}_{\text{processing time}} + \underbrace{\sum^{\mathcal{L}}_l\frac{1}{\mu_l}+\frac{\rho}{2\mu_l (1- \rho)} }_{\text{queuing per wired link}},
\end{equation}
where $\mathcal{L}$ is the series of links to be traversed to exchange the user context.
 We assume that the service cannot serve the user during user context migration, as it would update the context to be migrated.
%In \cite{liu_performance_2011}, $\rho$ for different services is presented. Once again, the dirty page rate change substantially from service to service and based on the workload, thus we consider the average $\rho\approx0.07$. We assume that while the migration of user-related memory pages afftects the service availability from the interested user perspective, it does not impact other users using the same service.
%
\subsection{Service Lifecycle}
\begin{figure*}[t]
    \centering
    \includegraphics[width=\textwidth]{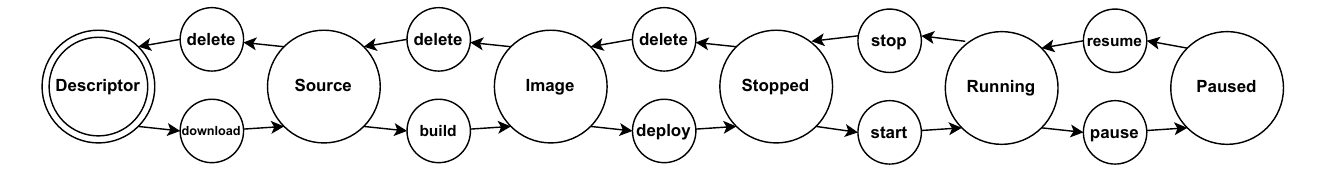}
    \caption{Lifecycle of a service}\label{fsm}
  \end{figure*}

\autoref{fsm} shows an \gls{fsm} that generalizes the different states through which a service passes before being available. These states are common in the vast majority of virtualized software, like \glspl{vm}\footnote{\url{https://docs.openstack.org/nova/latest/reference/vm-states.html}} and containers\cite{stahlbock_optimization_2022}, but can be easily extended to any kind of software.

In the descriptor state, only a descriptor of the service is available at the \gls{ec}, conventionally a configuration text file given by the service tenant. After retrieving the source of the service (e.g., source code, layers, modules), the service can be built. The built image can be deployed (i.e., creation of a \gls{vm} or container) and it is then in the stopped state. From there, it can be started and enters the \emph{running} state. 
Transitioning between service states requires time, but depending on the type of service to be run some of these transitions may be part of other processes or skipped. For instance, a \gls{vm} image is usually already built, but the deployment (allocation of resources, creation of a virtual disk, etc.) has to be performed. In this case, we can consider $t_{\text{build}}=0$.
\begin{table}[]
    \centering
    \caption{Resources usage at each state}
    \begin{tabular}{lcccc}
        \toprule
        State &  Disk & CPU & Memory  \\
        \midrule
        Descriptor  & x & x & x \\
        Source &  $\checkmark$ & x & x \\
        Image &  $\checkmark$ & x & x \\
        Stopped &  $\checkmark$ & x & x \\
        Paused    & $\checkmark$ & x & $\checkmark$\\
        Running &  $\checkmark$ & $\checkmark$ & $\checkmark$ \\
        
        \bottomrule
    \end{tabular}
    
    \label{tab:usage}
\end{table}
Different states require different types of resources, as shown in \autoref{tab:usage}. This can affect which and how many services can be deployed in which state when considering resource-constrained \glspl{ec}. 

We assume that the source and any file regarding the service can not be deleted during execution or without traversing the \gls{fsm} accordingly. 
Additionally, we assume that each \gls{ec} has always \emph{at least} the descriptor for each service. We argue that such descriptor, conventionally being a text file, does not impose any  burden on the storage of \glspl{ec}.

\section{Problem Formulation}
With the objective of meeting users' strict \gls{qos} requirements, we want to minimize the ratio of  users' \emph{unsuccessful packets}. 
A packet sent by users is considered unsuccessful if any of the following three conditions holds: 
  \begin{inparaenum}[(i)]
  \item The packet is lost anywhere, in particular, on the wireless link,\footnote{Packet losses are not evaluated in our example scenarios to maintain focus on key points, though they would surely render a packet unsuccessful.} or 
  \item when packet arrives at the EC serving the request VNF, that VNF is not in the running state, 
  \item or the VNF is running, but currently migrating user state (only relevant for stateful VNFs),
  \item or the running VNF processes the packet but fails to meet the deadline imposed by the service's \gls{qos} requirements.
  \end{inparaenum}
%-----
We use   $p_u(t)=1$ to denote that the packet sent by user $u$ at time $t$ is unsuccessful (and $p_u(t)=0$, else). 
\begin{subequations}
    \begin{equation}
        \min   \; \lim_{T\to\infty}\sum^{T}_t\sum^{\mathcal{U}}_u p_u(t),
    \end{equation}
%\end{subequations}
% \begin{subequations}
%     \begin{equation}
%         %\min\limits \lim_{T->\infty} \frac{1}{T|\mathcal{U}|} \sum^{T}_t \sum_u^\mathcal{U} m_u(t).
%         \begin{split}
%             \min \sum^\mathcal{U}_u\sum^\mathcal{E}_e |\mathcal{L}_{u,e}|\cdot x_{u,e} \\
%              + \sum^\mathcal{U}_u\sum^\mathcal{E}_e\sum^{\text{FSM}}_\psi z_{e,u_s,\psi} \cdot e_\psi^s,
%         \end{split}
%       \end{equation}

    \noindent Let $z^t_{e,v,\psi}$ the binary variable set to 1 if \gls{vnf} $v$ at \gls{ec} $e$ is in state $\psi$ at time $t$, and $c^t_{u,e}$ the binary variable set to 1 if user $u$ is connected to \gls{ec} $e$ at time $t$. We use $s_{e,v}(t)=1$ if \gls{vnf} $v$ in \gls{ec} $e$ is currently transitioning between states, $s_{e,v}(t)=0$ otherwise. Any solution to this problem must comply with these constraints: 
    \begin{equation}\label{c1}
        s_{e,v}(t)=1 \Rightarrow z^t_{e,v,\psi} =  z^{t-1}_{e,v,\psi} \quad \forall e\in\mathcal{B},v\in\mathcal{V},t\in T
    \end{equation}
    \begin{equation}\label{c2}
        \sum_{\psi}^\text{FSM} z^t_{e,v,\psi} = 1 \quad \forall e\in\mathcal{B},v\in\mathcal{V},t\in T
    \end{equation}
    
    \begin{equation}\label{c3}
        \sum^\mathcal{\mathcal{V}}_v R^e_{v} \leq R_{\text{tot}} \quad \forall e \in \mathcal{B},
    \end{equation}
    \begin{equation}\label{c4}
        \sum^\mathcal{\mathcal{B}}_e c^t_{u,e} =1 \quad \forall u \in \mathcal{U},\forall t \in \mathcal{T}.
    \end{equation}
    
  \end{subequations}

 Equation \eqref{c2} ensures that in each \gls{ec}, each service is in one and only one state at a time. Finally, \eqref{c3} ensures that resources at \glspl{ec} are not exceeded.
 %In an online problem, service placement in the network and the user connection to the deployed instances can be written as a \gls{ilp} problem and optimally solved for a specific timestep. However, with each state change requiring time to take effect, this \gls{ilp} solution is not guaranteed to be optimal due to user mobility during such changes highly affecting the optimal solution.
% With a perfect knowledge of the environment and its future, it would be possible to guarantee services to be in the desired state pre-emptively. But in reality, it is not possible to obtain such knowledge.

\section{Proposed approach}
We want to deploy \glspl{vnf} along users' paths to reduce unsuccessful packets but, due to their lifecycle, deployment must begin early to ensure readiness when needed. Moreover, given \glspl{ec}' constrained resources, over-provisioning of \glspl{vnf} is not possible. Consequently, \glspl{vnf} must be deployed in time and only where there is a reasonable chance they will be needed in the near future.

\subsection{Service deployment}
\label{sec:decid-serv-depl}

\begin{figure}
    \centering
    \includegraphics[width=\linewidth]{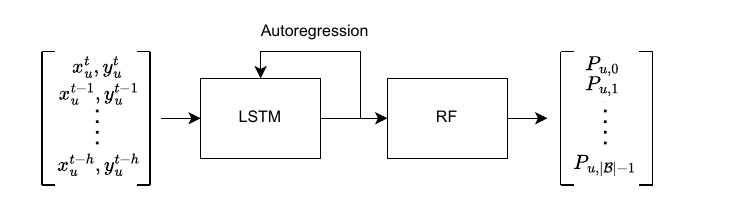}
    \caption{SWAVES inference chain.}
    \label{fig:forecasting}
\end{figure}
%To optimally change the services' states in each \gls{ec}, we need to know where the user could be\fmhkn{could be? meaning what? limit on travel speed?} connected in the next useful\fmhkn{define ``useful''?} time horizon $h$
%As users traverse cells, we want \glspl{vnf} to be running in cells that are likely to host users in the near future, to minimize unsuccessful packets.

We present our approach, SWAVES. By assessing the likelihood of a user connecting to different \glspl{bs}, we can change the required \glspl{vnf} state in the respective \glspl{ec} to \begin{inparaenum}[(i)]
    \item get \glspl{vnf} as close to the running state as possible where users will likely connect, and
    \item remove \glspl{vnf} where users are not going to connect, freeing \glspl{ec} for other \glspl{vnf} to use.
\end{inparaenum} 
While a \gls{vnf} near to the running state is not able to serve users, shortening the time to the running state can sensibly reduce unsuccessful packets.
%As a result, \glspl{vnf} will be available for user in high-likelihood nodes, or at most close to running, reducing unsuccessful packets. 
% \begin{figure}
%     \centering
%     \includegraphics[width=\linewidth]{img/user_moving.pdf}
%     \caption{As users traverse coverage areas, we want to prepare the service needed and remove it from where it is unnecessary.}
%     \label{fig:usermoving}
% \end{figure}
% Let $\mathcal{L}_u$ be the set of possible paths for user $u$, with $L \in \mathcal{L}_u$ being given as a sequence of locations $l_u(t)$, $L= < l(t), l(t+1), \ldots  >$, where $l(t)$ is the user's location at time $t$.\fmhkn{not sure it is necessary to use $l_u$; $l$ should be enough? } \fmhkn{actually, the following expostiion and notation would simplify a bit by limiting $\mathcal{L}$ to finite sets of paths. This is justified if
%   \begin{inparaenum}
%   \item we only look at finite horizons (of course)
%   \item use discrete time (plausible) 
%   \item use only discrete locations -- that would require some argumentation, but is probably defensible, too
% \end{inparaenum}. MAybe it is better to say ``finite'' here from the outset, rather than sneaking it in later when you introduce your LSTM and random forest prediction models; both are finite in nature, AFAICT.
% }
We assume knowledge of two probabilities (cmp. Section~\ref{sec:estim-prob}):
\begin{enumerate}
\item The probability density function $f(L)$ of a user following a particular path $L \in \mathcal{L}_u $
\item  The probability $P( b \,|\, l)$ of a user $u$  connecting to \gls{bs} $b$  given its location $l$. 
\end{enumerate}

From that, we compute the probability of the user \emph{not} connecting to a cell $b$ through the entirety of a time horizon $h$ by conditioning on which path $L$ a user actually follows, using the law of total probability, and then checking whether the user connects to $b$ at that particular location. In more detail: 
\begin{subequations}
    \begin{equation}\label{eq:swaves1}
        % P_{b_u\neq b} = \int_{\mathcal{L}_u}\prod_{i=1}^h\left[P(\neg \, b_u \,|\, l(t+i))\right]\cdot P(l)\; dl.
      P_{b_u\neq b} =
      \int_{L \in \mathcal{L}_u \atop L= < l(t), l(t+1), \ldots  > }
      \prod_{i=1}^h\left[ 1-P(b \,|\, l(t+ i)) \right]\cdot f(L)\; dL.
    \end{equation}
    
    To consider service multi-tenancy, given the \gls{vnf}  $v$ to deploy and  the subset $\mathcal{U}_v$ of users needing $v$, the probability of a service being needed at a certain cell $b$ is equal to 
    \begin{equation}\label{eq:swaves2}
        P_{v,b} = 1-\prod_{u \in \mathcal{U}_v} P_{b_u\neq b}
    \end{equation}
\end{subequations}
%\fmhkn{but that depends on the length of the prediction horizon, right? And for longer horizons with smaller probabilities, you could afford to put services in lower-effort states as the risk is lower. Basic SWAVES idea, really. Not sure that behvior would result from this prediction approqch? }
Computing  $P_{v,b}$ for each \gls{vnf}
% deployable in the network
gives us knowledge of where \glspl{vnf} are likely to be needed. We can exploit such knowledge during service deployment operations to prepare \glspl{vnf} in time, reducing users' unsuccessful packets.

\subsection{Estimating probablities}
\label{sec:estim-prob}

To
estimate the probabilities $P(b\,|\,l)$ and $f(L)$  needed in equation \eqref{eq:swaves1} in a real system, we employ the inference chain shown in \autoref{fig:forecasting}. Forecasting of user location is an autoregressive process where, given an input of the size of the horizon $h$ containing the last $h$ positions, the model outputs the next location. This process is repeated by sliding the input window and containing the previous cycle's output, resulting in a prediction of a path $L$.

%   ; this can be estimated in practice from actual user movement\fmhkn{provide reference here?} and is implicitly given by the mobility model in the simulations in Section~\ref{sec:numerical-evaluation}. 
% In practice, this can be estimated from \fmhkn{\dots not sure? }  and is easily available in simulations in Section~\ref{sec:numerical-evaluation} \fmhkn{I am not sure it is wise to foreshadow the simulations here,... two-edged sword. Needs more thinking.} \fmhkn{of course, this begs the question how you would know the location in real life? }

% This process's accuracy is affected by two factors: horizon length and randomness in the user movement. Increasing the horizon length decreases the forecasting accuracy, as the model uses more inferred data during predictions. Likewise, as users' movements become increasingly random, the model's prediction becomes less accurate. 

But users' connection to \glspl{bs} does not solely depend on their location: it is heavily influenced by randomness and other environmental factors. Consequently, we cannot expect to be exact in predicting to which \gls{bs} a user will connect given its position. Therefore, we use a classifier to output, given the user's location, the probabilities of one \gls{bs} to be a candidate for the handover. Finally, resulting probabilities are used in equations (\ref{eq:swaves1}) to predict the likelihood of a service being needed at a certain cell within the time horizon, influencing service placement decisions.

% Service location forecasts for the next time horizon are fed into an \gls{ilp} solver, to optimally decide both services' states in each \gls{ec} and users' connection to \glspl{ec}.

%The time horizon $h$ has to be chosen carefully: too long and the prediction quality inevitably degrades; too short and it could lead to erroneous placements. For this reason, we set the time horizon.

\section{Numerical Evaluation}
\subsection{Topology}
\begin{figure}
    \centering
    \includegraphics[width=\linewidth]{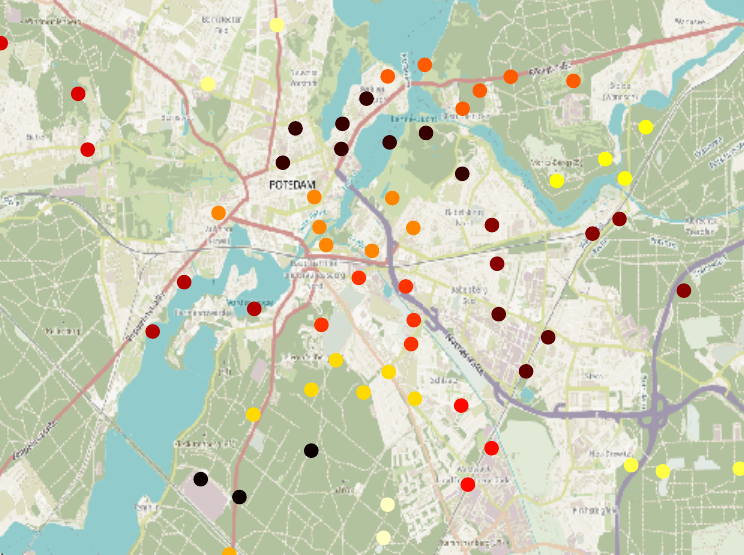}
    \caption{Distribution of \glspl{bs} in Potsdam and their clustering (by color).}
    \label{fig:bslocation}
\end{figure}
We consider \glspl{bs}' location inspired by Potsdam, Germany\footnote{Data extracted from cellmapper.net, manually refined to account for multiple vendors.}. \glspl{bs} are then connected with 
 a tree of multiplexing points. In particular, the root of the network, i.e., the $M_3$, is connected to 4 $M_2$, connecting all 16 $M_1$, as in \cite{ye_end--end_2019}. Finally, the 64 \glspl{bs} are split into 16 clusters given by proximity, as illustrated in \autoref{fig:bslocation}, each connected to an $M_1$.%Each cluster of \glspl{bs} is connected via a layer of multiplexing nodes (M1). M1s are connected by $4$ M2s and a single M3, considered the network's root.
%To maximize the benefit given from proximity of users and \glspl{ec}, we consider each \gls{bs} to have its own \gls{ec} directly connected and reasonably close. For this reason, from this point the term \gls{bs} also indicates the \gls{ec} connected to it.
\subsection{Mobility Model}
We chose the Gauss-Markov mobility model to describe user movements. In a 2d environment, the velocity $v(t)$ and direction $\Theta(t)$ of a user at time $t$ are determined as:
\begin{subequations}\label{mobility}
    \begin{equation}\label{mobility:vel}
        v(t) = \alpha v(t-1) + (1-\alpha)\bar{v}+\sqrt{(1-\alpha^2)}v_{\text{rnd}}
    \end{equation}
    \begin{equation}\label{mobility:dir}
        \Theta(t) = \alpha \Theta(t-1) + (1-\alpha)\bar{\Theta}+\sqrt{(1-\alpha^2)}\Theta_{\text{rnd}}
    \end{equation}
\end{subequations}
In (\ref{mobility:vel}) and (\ref{mobility:dir}), the values of velocity and direction at time $t$ are based on the same values at time $t-1$, their mean, and a random factor. The weight of randomness over historical values is determined by the parameter $\alpha$. While moving, users handover to new \glspl{bs} whenever their signal is better than the current \gls{bs} (LTE's A3 event) for a certain period, i.e., the \gls{ttr}.
%Though this mobility model works with discrete time, it is handled accordingly to other continuous time network simulators, like NS-3\footnote{\url{https://www.nsnam.org/}}, that update user position once every second.
%Handover schemes are not in the scope of this work, hence users are handed over to a new \gls{bs} whenever the latters' signal becomes better than the current \gls{bs} (LTE's A3 event). We also assume \gls{daps} to be enabled in order to avoid service disruptions from handovers \cite{lee_intelligent_2022}.  
\subsection{Latency computation}
Latency over a wireless link in a cellular system is not easy to simulate as it depends on, among others, link quality and competition for resources in a cell. We abstract from that and use the Shannon rate based on path loss to determine rate and thence time to transmit a packet.
We consider COST 231 Hata loss model \cite{akhpashev_cost_2016} to model path loss $L_p$. Using the link budget formula, we compute the power received by user $u$ from \gls{bs} $b$ as:
\begin{equation}
    P_u = P_b \frac{\abs{h}^2}{L_p},
\end{equation}
with $P_i$ the power received by user or transmitted by \gls{bs} $i$, $|h|^2$ the rayleigh fading and $L_p$ the path loss.
The data rate achievable by user $u$ can be computed as:
\begin{equation}
    \mu_u = B\: \log_2(1+\frac{P_u}{N}),
\end{equation}
with $B$ the channel bandwidth and $N$ the channel noise.
Finally, the total \gls{e2e} delay for users is computed as per \eqref{eq:delay}.
\subsection{Virtual Network Functions}
Lightweight virtualization of services is a common choice for deployment in \gls{mec}. Containers, as a lightweight virtualization technique, have gained popularity for their ease of deployment and flexibility. We consider containers to be available in a common repository. 
Fu \emph{et al.} \cite{fu_fast_nodate} present aggregated data about the top 5k container images on the public repository DockerHub. For example, the average pull time $\bar{t}_{\text{download}}=19.2 \;\text{s}$.
%Containers' file system, in its most popular implementation, is divided into layers. Layers can be shared among different images (e.g. when two images are based on the same Linux distribution), so there is no need to download locally-available layers multiple times. Though this can bring various benefits, such as decreasing the download time for container images, it requires containers to run the same underlying software at the same version. 
Docker avoids downloading layers already present from other containers, which can reduce $t_\text{download}$. However, since this relies on container-specific assumptions, we do not consider this aspect.

When using the default container manager for Docker, \texttt{containerd}, building an image (extracting and combining layers) is already a process of the image pull, so this time is already considered inside $t_{\text{download}}$. Hence, we assume $t_{\text{build}}=0$. 
%This is not necessarily true for other container managers, often seen in the literature to optimize the deployment process per se, but this is not the aim of this study. 
Yu \emph{et al.} \cite{yu_characterizing_2020} provide experimental absolute values for $t_{\text{deploy}}$ and $t_{\text{start}}$ (defined as the time to start the container process, not the service). Here, we can find quasi-constant values for $t_{\text{deploy}}\approx100\;\text{ms}$, $t_{\text{start}}\approx530\;\text{ms}$, $t_{\text{pause}}=t_{\text{resume}}\approx96\;\text{ms}$. 
In literature, no experimental values are found for  $t_{\text{stop}}$. Thus, we assume  $t_{\text{stop}}=t_{\text{start}}$ for a graceful stop of the \glspl{vnf}.
Finally, we consider any operation of deletion, i.e., any operation traversing the \gls{fsm} from the \emph{stopped} state to \emph{descriptor}, to take negligible time.

 We consider $R_v$ to be the same for all \glspl{vnf} and equal to (2,5,1): 2 CPU cores, 5 GB of RAM, and 1 GB of disk space. We also assume that containers' resource consumption is bounded by $R_v$, which remains constant regardless of user load. 
Finally, we consider that the available resources at \glspl{ec}, $R_{\text{tot}}$ is the vector of resources $(8,20,5)$.
% \begin{figure}
%     \centering
%     \includegraphics[width=\linewidth]{img/predictions.pdf}
%     \caption{SWAVES inference chain.}
%     \label{fig:forecasting}
% \end{figure}
\begin{figure}
    \centering
    \includegraphics[width=\linewidth]{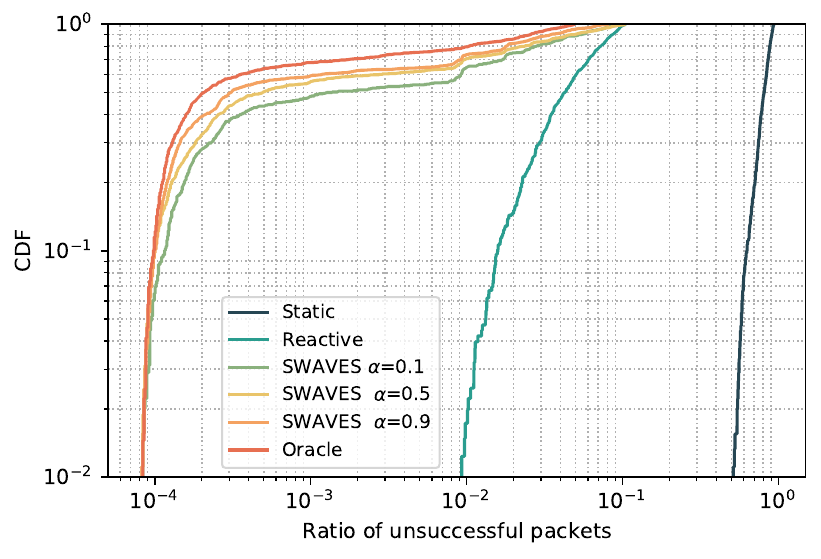}
    \caption{CDF of unsuccessful packets per user with a 1 ms delay limit.}
    \label{d1}
\end{figure}
\begin{figure}
    \centering
    \includegraphics[width=\linewidth]{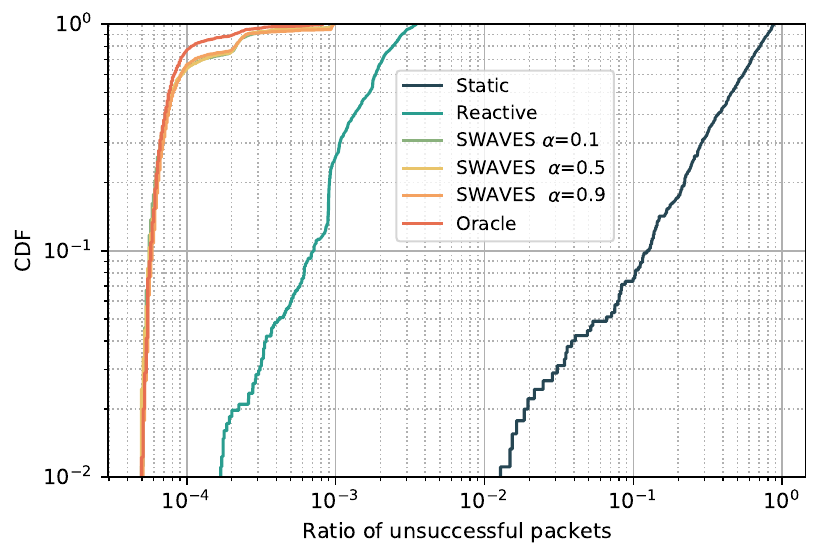}
    \caption{CDF of unsuccessful packets per user with a 2 ms delay limit.}
    \label{d2}
\end{figure}
\begin{figure*}
    \centering
    %\subfigure[Figure A]{\includegraphics[width=0.30\linewidth]{img/d2.pdf}\label{fig:d2b}}
    \subfigure[$\alpha=0.9$]{\includegraphics[width=0.45\linewidth]{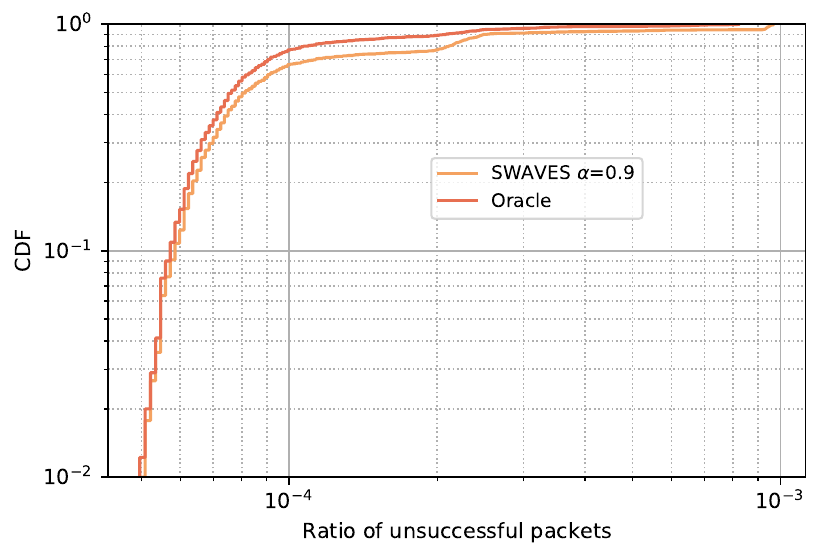}\label{fig:d201}}
    \subfigure[$\alpha=0.1$]{\includegraphics[width=0.45\linewidth]{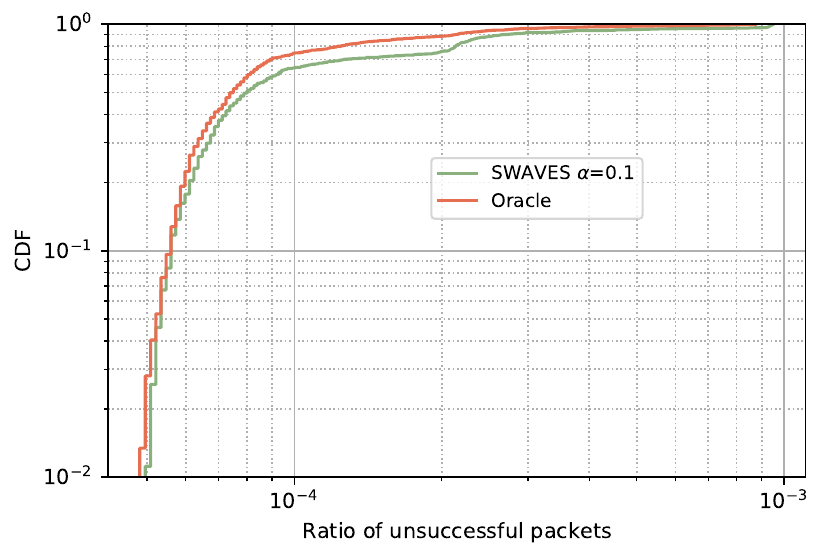}\label{fig:d209}}
    \caption{CDF of unsuccessful packets per user with a 2 ms delay limit for SWAVES and Oracle.}
    \label{fig:d2}
\end{figure*}

\begin{table}[]
    \centering
    \caption{Simulation parameters}
    \label{sim}
    \begin{tabular}{lc}
        \toprule
        Parameter & Value  \\
        \midrule
        $\text{Simulation duration}$ & 10 min \\
        $|\mathcal{U|}$ & 50 \\
        $h$ & 5 s\\
        \gls{e2e} delay limit & [1, 2] ms \\
        $\alpha$ & [0.1, 0.5, 0.9] \%\\
        $V_{\text{mem}}$ & 500 KB \\
        $\lambda_u$ & 0.2 Mbps \cite{firouzi_delay-sensitive_2024} \\
        $\mu_l$  & 1 Gbps \\
        $t_{p}$ & 0.2 ms \cite{coll-perales_end--end_2022}\\
        \bottomrule
    \end{tabular}
    
    \label{tab:param}
\end{table}
\subsection{Comparison cases}
We use the previously discussed modeling and experimental assumptions in a simulator. Parameters for the performed simulations can be found in \autoref{tab:param}.
We execute simulations of 10 minutes simulated time, during which 50 users move in the network using the same mobility model, uniformly requiring one of 10 available \glspl{vnf}.
We compare the SWAVES approach to three others:
\begin{inparaenum}[(i)]
    \item A static heuristic;
    \item A reactive heuristic;
    \item An oracle.
\end{inparaenum}
Literature on this topic has thoroughly explored the problem of service placement using a variety of methods and solvers. With the aim of demonstrating the impact our approach has on the provisioning itself, we perform \gls{vnf} placement and user connection decisions using an \gls{ilp} solver. As such, the approaches differ in \emph{when} the problems are solved, and whether future knowledge is used. 
%\fmhkn{also: ILP solver describes the how, but not the what? I think what you are trying to say is that you use a perfect solution for the placement problem, and to obtain that perfect solution, you use an ILP solver and let it solve the problem to completion. But it is not clear what is used as an input for this solver, nor is it clear what is left to do once the solver is done? When you think of the provisioining problem as something that decides where to run each service instance, that IS a placement problem. and when you let handover / wireless association to BS be solved by LTE standard handover approach, it sounds like there is nothing for your solver to do. That needs more explanation.  }

The \emph{static heuristic} solves the placement problem only once. In this case, no \gls{vnf} changes state, and no user is ever migrated between instances throughout the simulation. It is intended as almost trivial to realize, but likely a worst-case comparison. The \emph{reactive heuristic} solves the problem s every time an event takes place. Two types of events are used as triggers: a \gls{vnf} is ready or a handover happens. SWAVES works the same, but periodically perform forecasts for all users, and considers them for the placement and connections problems. Finally, \emph{the oracle} acts the same as SWAVES, but it  has perfect future knowledge.

These approaches are compared on the ratio of unsuccessful packets under different network conditions, like the randomness of user movement and the delay limit required by \glspl{vnf}. In the following sections, results for the described approaches are shown. To improve readability, each plot shows only one oracle line, obtained from simulations with $\alpha = 0.9$.
%In the next section, the performance of SWAVES and the presented baselines are reported with respect to different delay limits and randomness of user movement.

%In the next section, the performance of SWAVES and the presented baselines are reported with respect to different delay limits and randomness of user movement.

\subsection{Lowest delay limit case}

In this scenario, \glspl{vnf} bound  \gls{e2e} delay  to  1\, ms.
% strict \gls{qos} requirements, having to comply with a \gls{e2e} delay of 1 ms.
% To comply with such a low delay,
To support this, \glspl{vnf} \emph{must} be located where users are connected.
Due to the time \glspl{vnf} take to change state, it is paramount for the user to find the required \gls{vnf} already in a state that is as close as possible to running at the new \gls{ec}. \autoref{d1} shows a cumulative distribution function (CDF) of the ratio of unsuccessful packets per user.

The static heuristic shows that, in such conditions, up to 90\% packets can be unsuccessful. The reactive heuristic shows how constantly adapting the service location can enhance this metric by up to two orders of magnitude with respect to the static heuristic.
The reactive heuristic  not having any information of where the user may be headed results in setting up \glspl{vnf} in advanced enough states in nearby \glspl{ec}. This comes at the cost of filling \glspl{ec} with unused \glspl{vnf}, requiring time to shut them down and remove them to make space for useful ones; but these still need additional time to be started.

This is the main advantage of SWAVES. Thanks to forecasting, it is possible to use resources in high-likelihood \glspl{ec}, ensuring that \glspl{vnf} are prepared only where there is a reasonable chance for them to be useful.
Intuitively, SWAVES's performance is affected by the randomness of user movement. Still, the main gain of SWAVES is restricting where \glspl{vnf} are prepared, resulting in at most 15\% difference between SWAVES runs with different $\alpha$ , converging in the best and worst case. Likewise, the gap between SWAVES and the oracle is between 10\% and 20\%, depending on $\alpha$.
This gap is given by handovers being not deterministic, so even if SWAVES can predict the eligible \glspl{bs} for the users to connect to, the uncertainty results in more resource consumption from which the oracle does not suffer.

\subsection{Medium delay limit case}
In this scenario, the acceptable delay requirement is relaxed compared to the previous case. In particular, services can now tolerate a \gls{e2e} delay of up to 2 ms. Now \glspl{vnf} do not \emph{need} to be deployed at the exact location where users are connected. Conversely, user requests comply with such a delay limit when connected to \glspl{ec} that are part of the same $M_1$ cluster. \autoref{d2} shows the CDF of unsuccessful packets per user in this scenario.

Results for SWAVES, with respect to the value of $\alpha$, converge to very similar values.
% SWAVES, in particular, closes the optimality gap, converging with the oracle in 90$\%$ of the distribution, retaining a small gap in the worst cases. Here, SWAVES converges to the same values no matter the value of $\alpha$.
This is quickly explained: $M_1$ clusters are created by clustering groups of close \glspl{bs}, so even if the forecast is not accurate, it is enough to predict one of the \glspl{bs} of the same cluster. 
Additionally, the gap between SWAVES and the oracle decreases even more, presenting still a 10\% gap but for way fewer data points.
This can be seen in Figures \ref{fig:d201} and \ref{fig:d209}, that focus on the extremes of $\alpha$.
The distance between SWAVES and oracle can be attributed again to the randomness of the handover process, which could connect users to $M_1$ clusters different from the predicted ones.

\section{Conclusions \& Outlook}
Our approach, SWAVES, is able to leverage forecasts of user movement to prepare \glspl{vnf} at \emph{the right place and time}, sensibly reducing service disruption during their movement.
This is possible thanks to several factors. First, knowing likely future  users connection helps freeing resources and spending less time in \gls{vnf} deployment.
This allows users to fully exploit the proximity to the required \gls{vnf}, ensuring compliance with their strict \gls{qos} requirements. Consequently, users experience disruptions mostly only due to stateful migrations, reducing unsuccessful packets by orders of magnitude with respect to other heuristics.
Even when forecasts are inaccurate, SWAVES has been proven beneficial in enforcing service proximity to fewer cells, leading to an overall improvement concerning the reactive heuristic.
In this paper, the effectiveness of SWAVES in the context of single \gls{vnf} deployment and the importance of addressing the service lifecycle during service placement operations have been addressed. 
Future work on SWAVES will include the application to service chains and graphs, focusing on more complex service deployment scenarios.

\bibliographystyle{IEEEtran}
\bibliography{lib.bib}

\newpage

\end{document}